\documentclass[runningheads,a4paper]{llncs}

\usepackage{amssymb}
\usepackage{graphicx}

\usepackage{url}
\urldef{\mailsa}\path| mahyuddin@usu.ac.id|
\urldef{\mailsb}\path|yayank262000@yahoo.com|
\newcommand{\keywords}[1]{\par\addvspace\baselineskip
\noindent\keywordname\enspace\ignorespaces#1}

\begin{document}

\mainmatter  

\title{Terrorist Network: Towards An Analysis}

\titlerunning{Terrorist Network: Towards An Analysis}
 
\author{Mahyuddin K. M. Nasution$^{1)}$%
\thanks{Seminar International/Nasional Matematika dan Terapan (SiManTap2011), Medan, 28-29 November 2011}
and Maria Elfida$^{1)}$}

\authorrunning{M. K. M. Nasution \& M. Elfida}
\institute{$^{1)}$PS Sistem Informasi, STTH, Medan, Sumatera Utara, Indonesia\\
\mailsa\\
\mailsb\\}

\toctitle{SIMANTAP 2011}
\tocauthor{}
\maketitle

\begin{abstract}
Terrorist network is a paradigms to understand the terrorism. The terrorist involves a lot of people, and among them are involved as perpetrators, but on the contrary it is very difficult to know who they are caused by lack of information. Network structure is used to reveal other things about the terrorist beyond the ability of social sciences.
\end{abstract}

\section{Introduction}
The network is one of defining paradigms in modern era, or as concept for understanding the phenomenon in world \cite{nasution2010}. The most be noticed phenomenon is what we called as terrorism. As social phenomena, always, there is no single definition of terrorism that commands full international approval \cite{carlile2007}, but FBI define it as action of "the use of serious violence against persons or property, or threat to use such violence, to intimidate or coerce a government, the public or any section of the public, in order to promote political, social or ideological objectives" \cite{taylor2010}.  

In social network, the terrorism is a important study since 9/11 twin tower attack \cite{woods2011}. Social network analysis as knowledge extraction from graph is the study of mathematical models for interaction among persons, groups, or organizations \cite{mccallum2004}. The popularity of this study because the salience of the connections among terrorism actions. Thus, to find out the relations between actors of terrorism be important to do the counter of terrorism actions. Therefore, our goal is to describe some aspects of terrorist network. In the section of history: towards a motivation we discuss that historically the terrorist network is interaction of group people. Network Data and Investigation Section describes how to connect one with the other terrorist. Last section we describe a model of optimal network.

\section{History: Towards a Motivation}
Currently there is a modern issue, an intelligence agency is investigating general trends in terrorist activities all over the world. In the literature, initially emerged as a form of political expression, terrorism dates back to 6 A.D. when Jewish patriots opposed to Roman rule in Palestine, organized under the name of Zealots and launched a terrorist campaign to drive Romans out of Palestine \cite{poland1988,schlagheck1988}. Terrorism recurred from 116-117 A.D. and again from 132-135 A.D. until the Jewish population was driven out of Rome. The term did not officially enter political vocabulary until the 18th century, when Edmund Burke criticized the "reign of terror" following the 1792-94 French Revolution, when the French government used systematic terror to intimidate and eliminate its enemies \cite{murphy1989,poland1988}. On and off, the use of terrorism can be traced to presented day. In 20th century, the activity of international terrorism increased rapidly during the late 1960s and early 1970s, but after a brief quietness, the 1980s began and ended with terrorist violence until beginning of this century. By the end of the decade, terrorism has become commonplace \cite{damore1986,richter1986}. Comparatively fewer terrorist incident have been recorded for the first half of the 1990s. However, their nature and magnitude are not easily comparable to those of past years events as indicated by the US Department of State. Like that, terrorists in the modern era has always been associated with Jewish and Zionist. Therefore, the US Department of State defines terrorism as "... premeditated, politically motivated violence perpetrated against civilians and unarmed military personnel by subnational groups ... usually intended to influence an audience."; and international terrorism as "...involving citizens or the territory of more than one country" \cite{sonmez1998}. 

For example, as organization, Jemaah Islamiyah (JI) started as an Indonesian Islamist group and is a loosely structured organization characterized by territorial divisions: the peninsular Malaysia and Singapore; Java, Mindanao, Sabah, and Sulawesi; and Australia and Papua \cite{koschade2006}. Abdullah Sungkar motivated by the need for a new organisation that could work to achieve an Islamic State in Indonesia and cause of established JI in Malaysia around 1995. During the 1990's Al Qaeda infiltrated JI and JI subsequently developed into a pan-Asian network extending from Malaysia and Japan in the north to Australia in the south \cite{gunaratna2003}. Peak in Indonesia, the tactical operation of the Bali attack was conducted by JI's Indonesian cells until now the hunt terrorist operations continue take place. 

\begin{figure}
\centering
\includegraphics[height=7.78cm]{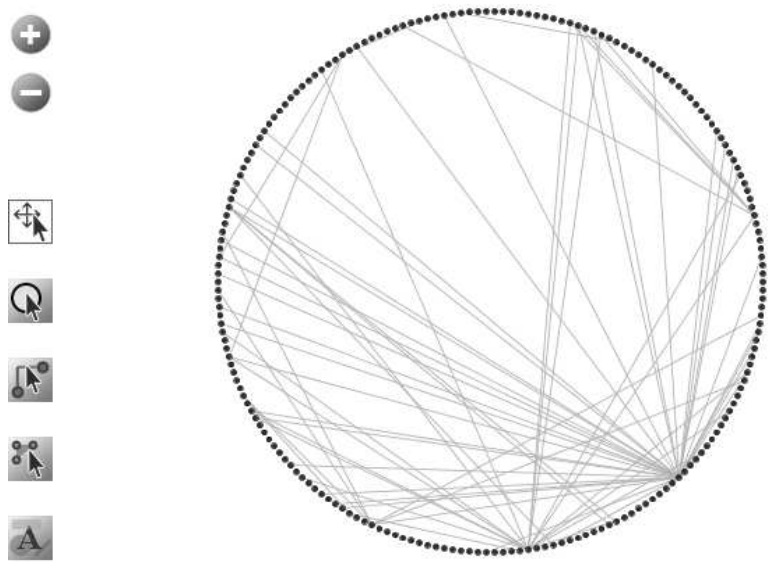}
\caption{The relations among entities in the salience}
\label{fig:concept}
\end{figure}

From histories, the actions of terrorism generally is not only done by a bunch of people but also the country to the people or the other country. Therefore, actions like this always to be a secret network. In modern era, the concept of the network allows us to capture the interactions of any individual unit within larger field of activity to which the unit exists, see Figure 1. The model of network based on the concept of graph. However, to construct any required network is the causes of the actors exist and related to the cluse of relations between them.\par

\section{Network Data and Investigation}
For building social network is difficult if only because a clue of relation between a pair of actors as a property of the pair, i.e. anything inherent to either actor. Collecting data on $n$ actors degenerates is to find the properties of $\frac{n(n-1)}{2}$ pairs of actors. Furthermore, the classical means of collecting such data by social scientists, though done carefully and reliably, are painstaking and time-consuming, is by involving questionnaires, interviews, direct observations, manual sifting through archival records, or various experiments \cite{wasserman1994}. 

Through investigation, there are two ways for finding the terrorist: a {\it direct} and an {\it indirect} ways. 
\begin{enumerate}
\item In particular, an independent Bernoulli random draw determines whether a particular agent is found {\it directly} as a terrorist. The direct finding about a particular agent as terrorist at each period is independent of other terrorists. Thus, a terrorist $i$ can be found directly by the authority according to some probability $\alpha_i$. This probability depends on the extent of the enforcement agency's scrutiny of terrorist $i$.
\item The law enforcement authority might also find terrorists indirectly. It is based on when the agency detects a terrorist who has information about other members of the organization, the agency also finds each of these with probability $\gamma \in (0,1]$. 
\begin{enumerate}
\item This implies that the enforcement agency's ability to find terrorists indirectly based on the structure social of the terror group. 
\item $\gamma$ is a parameter that depends on the ability of the law enforcement agency to extract information of terrorist activity from catched terrorists. 
\end{enumerate}
For instance, the parameter $\gamma$ is determined by the interrogation methods and the ability to strike by dealing with prisoners in exchange for information.
\end{enumerate}

The law enforcement agency has a budget $B\in [0,1]$ to allocate the resources for investigating the $N$ members of the organization and devoted $\alpha_i\in [0,1]$ to investigate member $i$ where $\sum_{i=1}^N\alpha_i\leq B$. Without loss of the generalization, we label terrorists so that $\alpha_1\leq\alpha_2\leq\dots\alpha_N$. We refer to $\alpha_i$ as the enforcement agency's level of scrutiny (or investigation) of terrorist $i$. Each terrorist can be found by the enforcement agency. If the terrorist is found, this imposes a direct cost to the organization of $k > 0$. This cost may include a punishment for the individual, such as time in prison, and a cost to the organization of recruiting a new member. 

In information structure, it is not just what you know, but also whom you know, that matters. What you know, say, $\Sigma$, $\Sigma\subset\Gamma$ the set of all thinking. Someone has only $mod(\Sigma)$ as a part of the thinking or class of knowledge $K$ \cite{nasution2010b}.

\begin{definition}
Let $\sigma$ be a generator as a trigger of the thinking, and $\{\sigma_i|i=1,\dots,m\}$ is a set of generator for $\Sigma$ such that $\Sigma = \{f(\sigma_i)|i=1,\dots,m\}$ is the knowledge of an actor where $f()$ as disseminator.
\end{definition}

A graph $G$ is an ordered pair $(V,E)$, where $V\ne\emptyset$ represents the finite set of verteces and $E$ represents the set of edges as set of all unordered pairs of vertices. The set of all graphs of order $n$ and size $m$ is denoted with $G(V,E)=G(n,m)$.  Let us define $\xi$ and $\zeta$ as two mapping, $\xi : A\rightarrow V$ and $\zeta : \{\Sigma_{a_i}\sqcap\Sigma_{a_j}\}_{i,j}\rightarrow E$, we obtain
a tie between $a_i$ and $b_j$ as edge 
\begin{equation}
\{i,j\} =\zeta(|\sigma_{a_i}\sqcap\sigma_{a_j}|)
\end{equation}
or briefly $ij=\zeta_{ij}$.

\begin{definition}
The shortest distance between vertices $i, j\in G$ is called the geodesic distance between $i$ and $j$, denoted by $\ell_{ij}(G)$, i.e., 
\begin{equation}
\ell_{ij}(G) = \min(\sum_{ij\in E}\zeta_{ij})
\end{equation}
where
\begin{enumerate}
\item $\ell_{ij} = \ell_{ji}$ if $ij = ji$, for all $ij,ji\in E$ or $ij$ is an undirected edge (symmmetry).
\item $\ell_{ij}\ne\ell_{ji}$ if $ij$ is a directed edge (assymmetric).
\end{enumerate}
\end{definition}

\begin{definition}
Let $G(V,E)$ is a graph, where $V\ne\emptyset$. The total distance, $T(G)$, is
\begin{equation}
T(G) = \sum_{i\in G}\sum_{j\in V}\ell_{ij}(G)
\end{equation}
\end{definition}

\begin{definition}
A diameter of a graph $G(V,E)$ defined as a maximum over the geodesic distances between all pairs of vertices, i.e., 
\begin{equation}
D(G) = \max_{ij\in V\times V}\ell_{ij}(G)
\end{equation}
\end{definition}

\begin{definition}
The community (based on fields of knowldge, or organization) is a set neighbors at distance $\delta$ of vertex $i$ by $\Xi_{i,\delta}(G)$, i.e., 
\begin{equation}
\Xi_{i,\delta}(G) = \{i\in V|\ell_{ij}(G)=\delta\}.
\end{equation}
\end{definition}

Some researchers and goverment agencies gather information from various sources that involved computer network or the computer mediated communication. They have a database of millions of new feed, minutes and e-mails and want to use these to get a detailed overview of all terrorist events in a particular geographic region in the last five years. This information not only related to the general public but also the information coming from government officials. Knowledge of terrorist networks \cite{krebs2002} is useful in finding various other crimes. In law enforcement concerning organized crimes such as drugs and money leundering \cite{xu2004}. Knowing patterns of relationship in a social network is very useful for law enforcement agencies to investigate collaborations among criminals, i.e., how the perpetrators are connected to one another would assist the effort to disrupt a criminal act or to identify additional suspects. 

\section{Optimal Network}
Any organization takes their network structure explicitly. In absence of further information, we are interested in what structure these organizations actually adopt. Theoretically, the analysis of the sharing knowledge of covert networks depends on proposed framework \cite{lindelauf2008} such that the optimal network structure derived the appropriate scenarios.

\begin{definition}
Let $G^n$ is a set of graphs of order $n$. The information measure of $g\in G^n$ is given by
\begin{equation}
K(g) = \frac{n(n-1)}{T(G)}
\end{equation}
\end{definition}

In terrorist organizations case, for two agents, one responsible for network secrecy and the other one for information efficiency, the set $G^n$ is connected graphs where the bargaining as an event in time space. Then, the secrecy as the hidden knowledge, we used $mod(\Sigma_{a_i})$ as a part of the class of knowledge.

\begin{definition}
The hidden knowledge measurement $H(g)$ of a graph $g\in G^n$ is defined as the expected fraction of the network that remains hid under assumption of sharing probability of individual $i\in V$ being equal to $\partial_i$, i.e., $mod(\Sigma_{a_i})$, as following
\begin{equation}
H(g) = \sum_{i\in V}\partial_iu_i
\end{equation}
where fraction of the network that individual $i$ sharing be $1-u_i$.
\end{definition}
$mod(\Sigma_{a_i})$ is uncountable, then $H(g)$ has one of tasks to cover it implicitly.
\begin{definition}
The balance among hidden and sharing of knowledge is an optimal graph $g\in G^n$ that maximizes
\begin{equation}
\gamma(g) = H(g)K(g)
\end{equation}
\end{definition}

The optimal graph is the bargaining solution of $g\in G^n$, and the probability of exposure of an individual in the organization is uniform over all network members, i.e, $\alpha_i=\frac{1}{n}$. The fraction of the network that individual $i$ exposes is 
\begin{equation}
1-u_i =\frac{pd_i+1}{n}
\end{equation}
where $p$ is probability identically if communication on links is detected independently, for $d_i$ as degree of vertex $i$ in graph $g$, i.e. if $p$ is a low value the complete graph is optimal. This statement equivalence to a lemma as follows.
\begin{lemma}
If $p\in[0,\frac{1}{2}]$, then $\mu(g_{comp}^n)\geq\mu(g), \forall g\in G^n$, where $g_{comp}$ is a complete graph on $n$ vertices.
\end{lemma}
And if $p$ is a high value the star graph is optimal.
\begin{lemma}
if $p\in[\frac{1}{2},1]$, then $\mu(g_{star}^n)\geq\mu(g), \forall g\in G^n$, where $g_{comp}$ is a star graph on $n$ vertices.
\end{lemma}

The actual network of operation of JI Bali bombing has been provided \cite{koschade2006}. This network is graph that we use as basis for comparison with the theoretical framework presented earlier. The Bali Bombing cell can be split into the bom making team, the support team and the command team. Both cells adopted the structure of a complete graph and obtained the optimal graph according to the theoretical framework. The command team visited both cells and coordinated the operation. Therefore, the characteristics of the sharing knowledge, ideology of terrorism, information, etc. for the number of linked terrorists $n$ is strictly less than $N$ if we take the authority's scrutiny $\{\alpha_1,\dots,\alpha_N\}$ and we study the most efficient information structure that the $N$ terrorists can form, as follows \cite{baccara20008,farley2006}.
\begin{enumerate}
\item The optimal structure is used to link $n<N$ terrorists is a hierarchy consist of the terrorist with the lowest probability of finding at the top, and the $n$ terrorists with the highest probability of the finding linked to him, i.e., $N, N-1,\dots,N-n+1\rightarrow 1$.
\item If $\gamma=1$, the optimal information structure is as follows:
\begin{enumerate}
\item If $N^* = 0$, the optimal information structure is an anarchy.
\item If $0<n^*<N$, the optimal structure is an individual-dominated hierarchy for where the hub is terrorist 1 and the subordinates are terrorist $N,\dots, N-n+1$.
\item If $n^*=N$, the optimal structure is the mixed structure.
\end{enumerate}
\end{enumerate}

\section{Conclusion}
Organisation of terrorists is a group of actor in the network structure,  where each group is connected efficiently with other groups to protect them. This connection forms can be examined using graph theory by considering all the information that it might exist.


\begin{thebibliography}{100}
\bibitem{nasution2010} Nasution, M. K. M. and Noah, S. A. 2010. Extracting social networks from Web documents. {\it The 1st National Doctoral Seminar on Artificial Intelligence Technology} (CAIT2010), UKM: 278-281.
\bibitem{carlile2007} Carlile, L. 2007. The definition of terrorism. A report by lord Cartile of Berriew Q.C. Independent Reviewer of Terrorism Legislation. {\it Cm 7052}. London: HMSO.
\bibitem{taylor2010} Taylor, M. 2010. Is terrorism a group phenomenon? {\it Aggression and Violent Behavior}, 15: 121-129.
\bibitem{woods2011} Woods, J. 2011. The 9/11 effect: Toward a social science of the terrorist threat. {\it The Social Science Journal}, 48: 213-233.
\bibitem{mccallum2004} McCallum, A., Corrada-Emmanuel, A., and Wang, X. 2004. The author-recipient-topic model for topic and role discovery in social networks: Experiments with Enron and Academic Email. {\it Technical Report} UM-CS-2004-096.
\bibitem{poland1988} Poland, J. M. 1988. Understanding Terrorism. Englewood Cliffs NJ: Prentice-Hall.
\bibitem{schlagheck1988} Sclagheck, D. M. 1988. International Terrorism. Lexington MA: Lexington Books.
\bibitem{murphy1989} Murphy, J. F. 1989. State Support of International Terrorism. San Francisco: Westview Press.
\bibitem{damore1986} D'Amore, L. J., and T. E. Anuza. 1986. International Terrorism: Implication and Challenge for Global Tourism. {\it Business Quarterly}, November: 20-29.
\bibitem{richter1986} Richter, L. K. and W. L. Waugh, Jr. 1986. Terrorism and Tourism as Logical Companions. {\it Tourism Management} 7: 230-238.
\bibitem{sonmez1998} S\"onmez, S. F. 1998. Tourism, terrorism, and political instability. {\it Annals of Tourism Research}, 25(2): 416-456.
\bibitem{koschade2006} Koschade, S. 2006. A social network analysis of Jemaah Islamiyah: The application to counterterrorism and intelligence. {\it Terrorism and Policical Violence}, 29: 559-575.
\bibitem{gunaratna2003} Gunaratna, R. 2003.{\it Inside Al Qaeda: Global Network of Terror}. Berkley Trade.
\bibitem{wasserman1994} Wasserman, S., and Faust, K. 1994. {\it Social Network Analysis: Methods and Applications}, Cambridge University Press.
\bibitem{nasution2010b} Nasution, M. K. M. {\it to appear}. Theory of sharing knowledge: an introduction. {\it Bulletin of Mathematics}.
\bibitem{krebs2002} Krebs, V. E. 2002. Mapping networks of terrorist cells. {\it Connections}, 24: 43-52.
\bibitem{xu2004} Xu, J., and Chen, H. 2004. Fighting organized crimes: using shortest-path algorithms to identify associations in criminal networks. {\it Decision Support System}, 38: 473-487.
\bibitem{lindelauf2008} Lindelauf, R. H. A., Borm, P. and Hamers, H. 2008. The influence of secrecy on the communication structure of covert networks. {\it CentER Discussion Paper}, 2008-23: 1-18.
\bibitem{baccara20008} Baccara, M., and Bar-Isaac, H. 2008. How to organize crime. {\it The Review of Economic Studies}, 75: 1036-1067.
\bibitem{farley2006} Farley, J. D. 2006. Breaking Al Qaeda Cells: A mathematical analysis of counterterrorism operations (a guide fr risk assessment and decision making). {\it Studies in Conflict and Terrorism}, 26: 399-411.
\end{thebibliography}
\end{document}